\documentclass[tradiabstract]{aa}
\usepackage{graphicx}
\usepackage{amsbsy}
\usepackage{natbib}
\usepackage{color}
\usepackage{txfonts}
\usepackage{url}
\usepackage{bm}

\DeclareMathSymbol{\varOmega}{\mathord}{letters}{"0A}
\DeclareMathSymbol{\varSigma}{\mathord}{letters}{"06}
\DeclareMathSymbol{\varPsi}{\mathord}{letters}{"09}
\DeclareMathSymbol{\varPhi}{\mathord}{letters}{"08}
\DeclareMathSymbol{\varGamma}{\mathord}{letters}{"00}

\begin{document}

\title{Exploring the conditions for forming cold gas giants\\ through
planetesimal accretion}
\titlerunning{Forming cold gas giants through planetesimal accretion}

\author{Anders Johansen\inst{1} and Bertram Bitsch\inst{2}}
\authorrunning{Johansen \& Bitsch}

\institute{$^1$ Lund Observatory, Department of Astronomy and Theoretical
Physics, Lund University, Box 43, 221 00 Lund, Sweden, \\e-mail:
\url{anders@astro.lu.se} \\ $^2$ Max-Planck-Institut f\"ur Astronomie,
K\"onigstuhl 17, 69117 Heidelberg, Germany \\e-mail: \url{bitsch@mpia.de}}

\date{}

\abstract{The formation of cold gas giants similar to Jupiter and Saturn in
orbit and mass is a great challenge for planetesimal-driven core accretion
models because the core growth rates far from the star are low. Here we model
the growth and migration of single protoplanets that accrete planetesimals and
gas.  We integrated the core growth rate using fits in the literature to
$N$-body simulations, which provide the efficiency of accreting the
planetesimals that a protoplanet migrates through. We take into account three
constraints from the solar system and from protoplanetary discs: (1) the masses
of the terrestrial planets and the comet reservoirs in Neptune's scattered disc
and the Oort cloud are consistent with a primordial planetesimal population of a
few Earth masses per AU, (2) evidence from the asteroid belt and the Kuiper belt
indicates that the characteristic planetesimal diameter is 100 km, and (3)
observations of protoplanetary discs indicate that the dust is stirred by weak
turbulence; this gas turbulence also excites the inclinations of planetesimals.
Our nominal model built on these constraints results in maximum protoplanet
masses of $0.1$ Earth masses. Ignoring constraint (1) above, we show that even a
planetesimal population of 1,000 Earth masses, corresponding to 50 Earth masses
per AU, fails to produce cold gas giants (although it successfully forms hot and
warm gas giants). We conclude that a massive planetesimal reservoir is in itself
insufficient to produce cold gas giants. The formation of cold gas giants by
planetesimal accretion additionally requires that planetesimals are small and
that the turbulent stirring is very weak, thereby violating all three above
constraints.}


\maketitle

\section{Introduction}

The catalogue of exoplanets is growing continuously. The observed planets
can be broadly divided into an inner population of planets, orbiting within a
few astronomical units (AU) from the star, and an outer population on more
distant orbits \citep[see][for a recent review of the exoplanet
populations]{WinnFabrycky2015}.  Planets from the inner population, that is, hot and
warm gas giants, super-Earths, and terrestrial planets, are relatively
straightforward to form in computer simulations, because the rates of
accreting planetesimals and pebbles so close to the star are high
\citep{ColemanNelson2014,Lambrechts+etal2019}, and because the inner edge
of the protoplanetary disc can stop migrating trains of protoplanets
\citep{Masset+etal2006,CresswellNelson2006}.

Planets in the outer population are far more challenging for planet formation
theories because growth rates are lower there and no general
migration barriers exist so far away from the edge of the protoplanetary disc. While
gas-giant planets similar to Jupiter in our solar system orbit 15\% of
solar-type stars out to a few AU from the star \citep{Mayor+etal2011}, the
occurrence rate drops to 3-6\% on $\sim$5 AU orbits \citep{Rowan+etal2016}.
Direct-imaging surveys find occurrence rates in the range of 1-6\% on orbits beyond
20 AU \citep{Vigan+etal2017}. 

Gas giants in cold orbits (which we define here to be broadly the regions
beyond 2--3 AU) are therefore generally rare, but a significant fraction of
the giant planets found in the inner population (hot and warm gas giants) were
likely scattered there from their primordial cold orbits \citep[][see also the
review by \citealp{Davies+etal2014}]{DawsonMurray-Clay2013,Buchhave+etal2018}.
The overall occurrence rate of gas giants increases rapidly with the metallicity
of the star \citep{UdrySantos2007,Buchhave+etal2012}, which make cold gas giants
excellent probes of the effect of the planetary growth rate (which is proportional to the
metallicity) on the outcome of planet formation. Cold gas giants also pose
the strongest necessity for rapid planetary growth because growth timescales of
more than a few million years will preclude the accretion of gas onto the core.

In this paper we study core formation by accretion of planetesimals. This is the
classical mode of core accretion \citep{Pollack+etal1996}, although many studies
employing $N$-body methods have questioned whether planetesimal accretion
is able to form the cores of cold gas giants before the
gaseous protoplanetary disc dissipates
\citep{Thommes+etal2003,Levison+etal2010,ColemanNelson2014}. Here we formulate a
similar concern based on studying the growth tracks of single migrating
protoplanets. We base our models on three constraints from the solar system and
protoplanetary discs for (1) the total planetesimal mass, (2) the characteristic
planetesimal size, and (3) the strength of the inclination stirring by the
turbulent gas.

The formation environment of exoplanets, in terms of the diameters and total
mass of the primordial planetesimal populations that contributed to core
accretion, is relatively poorly constrained. Observations of very young stars in
the embedded class 0 phase \citep[e.g.][]{Tychoniec+etal2018} show that the
protoplanetary discs in this phase are often very massive (containing up to
several tens of percent of the mass of the central star) and compact (from a few
ten AU down to a few AU in size). These young stars also display high accretion
rates in the range between $10^{-7}$ $M_\odot\,{\rm yr}^{-1}$ to $10^{-6}$
$M_\odot\,{\rm yr}^{-1}$ \citep{Hartmann+etal2016}. These masses and accretion
rates are consistent with models of the formation of a protostar and a
surrounding protoplanetary disc by the collapse of a giant molecular cloud core.
The angular momentum transport and mass accretion is driven by gravitational
instabilities and spiral arms in these earliest phases of star formation and
disc evolution \citep{Dullemond+etal2006}. If planetesimals form very
efficiently during these earliest stages of protoplanetary disc evolution, then
young protoplanetary discs could form several hundred or even a few thousand
Earth masses ($M_{\rm E}$) of planetesimals. A gravitationally unstable disc
converting 1\% of its gas surface density into planetesimals would produce
approximately 50 $M_{\rm E}$ of planetesimals per AU. Alternatively, pebble
pile-ups in the inner regions of the disc could produce very large amounts of
planetesimals over timescales of millions of years
\citep{DrazkowskaDullemond2018} if a significant fraction of the
inwards-drifting pebbles are converted into planetesimals.

The remnant planetesimal populations in the solar system nevertheless provide us with
a possibility of probing the properties of the primordial planetesimals that
formed around the young Sun.  Terrestrial planet formation simulations
successfully form Earth and Venus analogues starting from protoplanets with a
total mass of a few $M_{\rm E}$ \citep{Raymond+etal2009,IzidoroRaymond2018},
more than an order of magnitude lower than the above scenario of a very massive
planetesimal population would predict. Adding more mass than this to the
terrestrial planet zone instead results in the formation of a population of
migrating super-Earths \citep{Ogihara+etal2015}.

A part of the primordial planetesimal population beyond Neptune now resides in
Neptune's scattered disc and in the Oort cloud.  \cite{Brasser2008} found that a
primordial planetesimal population of approximately 35 $M_{\rm E}$ outside of
Neptune is consistent with the inferred mass of the outer Oort cloud ($\sim$$1
M_{\rm E}$).  This value, a few $M_{\rm E}$ per AU,  is again at least an
order of magnitude below what would be produced if planetesimals formed at the
self-gravitating stage of the protoplanetary disc.

The planetesimal mass per AU in the 5--10 AU region could in principle have been
one to two orders of magnitude higher than what is inferred from interior and
exterior regions, perhaps as the result of planetesimal formation by pebble
pile-ups exterior of the water ice line \citep{DrazkowskaDullemond2018}.
However, simulations of the early migration history of Jupiter show that
$\sim$$1\,M_{\rm E}$ of planetesimals is scattered in front of Jupiter and
parked in the asteroid belt \citep{RaymondIzidoro2017,Pirani+etal2019}, when a standard planetesimal population comparable to that of the minimum
mass solar nebula is assumed \citep{Hayashi1981}. Scaling these models to a much more
massive planetesimal population would imply injection of 50 $M_{\rm E}$ or more
of planetesimals into the asteroid belt, which would raise the question of how
this enormous amount of planetesimals could subsequently have been removed
\citep{Petit+etal2001}.

The birth sizes of planetesimals are also well constrained in the solar system.
The size distribution of asteroids in the main belt is consistent with formation
of bodies of a characteristic size of 100 km
\citep{Bottke+etal2005,Morbidelli+etal2009} and subsequent growth of larger
bodies (up to Ceres mass or even to the mass of protoplanets) by accretion of
millimeter-sized pebbles \citep{Johansen+etal2015}. The shallow size
distribution below the characteristic asteroid size, top-heavy with the
differential number of asteroids per mass unit ${\rm d}N/{\rm d}M \propto
M^{-1.6}$, is consistent with the initial mass function of planetesimals
produced by the streaming instability
\citep{Johansen+etal2015,Simon+etal2016,Abod+etal2018}, but could also arise
from accretion of planetesimals of initially 100 m in size
\citep{Weidenschilling2011}. A general pathway to form such small planetesimals
is still unknown \citep{Johansen+etal2014}. The size distribution between
10 and 100 km of cold classical Kuiper belt objects has a similar shape to
planetesimals formed by the streaming instability, followed by a shallow
exponential tapering at sizes above 100 km \citep{Schafer+etal2017}. The absence
of craters from impactors of sizes below 1 km on Pluto, Charon, and the cold
classical Kuiper belt object MU69 \citep{Singer+etal2019,Stern+etal2019} and the
high binary fraction of cold classical Kuiper belt objects
\citep{Nesvorny+etal2019} give further evidence of the formation of large
planetesimals from gravitationally bound pebble clumps
\citep{Johansen+etal2014}.

The inclination of the planetesimal population is another important parameter in
setting the efficiency of planetesimal accretion. Planetesimal inclinations are
mainly stirred by mutual scattering and by gravitational torques from the
turbulent gas \citep{Ida+etal2008}. Recent measurements of the scale height of
the dust layer in protoplanetary discs are consistent with a weak turbulence
with a diffusion coefficient between $10^{-4}$ and $10^{-3}$
\citep{Pinte+etal2016,Dullemond+etal2018}. We show that even these levels of
turbulence stir the planetesimals to inclinations of about $0.01$,
which reduces the core growth rate significantly compared to models neglecting
turbulent stirring.

Using these combined constraints in our model for core growth by planetesimal
accretion, we demonstrate that protoplanets reach a maximum mass of $0.1\,M_{\rm
E}$ after three million years of disc evolution. If we ignore the constraints
from the solar system on the total planetesimal population and allow
planetesimals to form at the self-gravitating stage, yielding a total
planetesimal mass of nearly 1,000 $M_{\rm E}$ (approximately 50 $M_{\rm E}$
per AU), then our model produces hot and warm gas giants, but no cold
objects similar to Jupiter and Saturn in mass and orbit. We show that we must
additionally lower the characteristic planetesimal size as well as the turbulent
stirring level by a factor 10 each to be able to form cold gas giants.

The paper is organised as follows. In Section \ref{s:core_growth} we review
numerical studies of core growth by planetesimal accretion. Particularly, we
focus on the large literature that exists on $N$-body simulations demonstrating
that the formation of Jupiter's core is slowed down significantly by gap
formation in the planetesimal population and by the oligarchic growth of
multiple bodies competing for planetesimals. In Section \ref{s:equations} we
present the equations that we solve in order to produce growth tracks of
protoplanets that migrate and accrete planetesimals and gas in an evolving
protoplanetary disc. The results are presented in Section \ref{s:growth}. Our
main finding is that a planetesimal population of 1,000 $M_{\rm E}$ is
insufficient to form cold gas giants and that the planetesimal sizes and the
turbulent stirring must additionally be significantly lower than their
nominal values in order to form analogues of Jupiter and Saturn. We summarise
and discuss our results in Section \ref{s:summary}. In Appendix A we
present a pebble-driven planet formation model, based on constraints from
observations of pebbles in protoplanetary discs and experiments on dust
coagulation, to give perspective to the planetesimal-driven planet formation
model considered in the main paper.

\section{Core growth with planetesimals}
\label{s:core_growth}

We start by reviewing in this section the literature that exists on the
formation of cold gas giants, similar to Jupiter in mass and orbit, by planetesimal
accretion. We focus on papers that concern $N$-body simulations because they
capture the radial transport of planetesimals in connection with\ gap
formation in the planetesimal population, for instance, and the oligarchic growth of
multiple protoplanets.

Simulations of core formation can broadly be categorised into four quadrants in
the space spanned by single or multiple in one direction and
migrating or non-migrating along the other: (1) single non-migrating protoplanets,
(2) single migrating protoplanets, (3) multiple non-migrating protoplanets,
and (4) multiple migrating protoplanets. We review the literature on these four
quadrants in the next subsections.

\subsection{Single non-migrating protoplanets}

\cite{TanakaIda1997} simulated the dynamics of a single protoplanet in an ocean
of smaller planetesimals. They observed that the planetesimals scattered by the
protoplanet circularise by gas drag beyond the reach of the protoplanet. The
growth of the protoplanet during gap formation was explored in
\cite{Rafikov2001}. Using Eq. (25) of \cite{Rafikov2001}, we calculate that
a gap is opened when the protoplanet has achieved a mass roughly between that of the
Moon and that of Mars in the 5--10 AU regions. Very small planetesimals are able to
penetrate into the gap by radial drift, but they must be smaller than a few
kilometers to drift fast enough to penetrate. A phase of rapid gas
accretion would also allow the core to access the small planetesimals trapped at
the gap edges \citep{ShiraishiIda2008,ZhouLin2007}, but such rapid growth is
generally not obtained during core formation, and particularly not at the early
stages of core growth. 

\subsection{Multiple non-migrating protoplanets}

Already \cite{TanakaIda1997} suggested that the problem of gap formation could
be alleviated by the simultaneous growth of multiple protoplanets.  Terrestrial
planet formation simulations show that such an oligarchic growth phase is a
robust outcome of planetesimal accretion \citep{KokuboIda2000}.  Oligarchic
growth in the giant planet region was simulated by \cite{Thommes+etal2003},
neglecting planetary migration from the gas torque.  The oligarchic growth shows
no strong gap formation. Nevertheless, \cite{Thommes+etal2003} required a
surface density of 200 kg/m$^2$ in planetesimals of 10 km in size at 5 AU
(translating into approximately 24 $M_{\rm E}$ per AU) to form cores of
$5\,M_{\rm E}$ there. Modern solar abundances yield a total condensible solids
fraction of $Z \approx 1.5\%$ \citep{Lodders2003}, but a large fraction of the
oxygen will be bound in supervolatile CO rather than water
\citep{Bitsch+etal2019} and hence will not contribute to the planetesimal mass
outside of the water-ice line.  With conservative value for the relative mass
bound in planetesimals of $Z_{\rm pla} \approx 0.5$, the successful models of
\cite{Thommes+etal2003} require the gas surface density of a gravitationally
unstable disc with self-gravity parameter $Q \approx 1$.
\cite{Levison+etal2010} instead considered the growth of four initially massive
protoplanets, but observed that these simply scattered a joint gap in the
population of planetesimals of 100 km in size. This result highlights that
placing protoplanets within a limited spatial interval can strongly reduce the
mutual feeding of the protoplanets in the oligarchic growth picture and instead
leads to the transport of planetesimals to beyond the reach of the protoplanets.

\subsection{Single migrating protoplanets}

The accretion rate of a migrating protoplanet was considered by
\cite{WardHahn1995} and by \cite{TanakaIda1999}. At slow migration (and/or small
planetesimals experiencing short gas drag timescales), the protoplanet becomes a
shepherd as the planetesimals are pushed in front of the protoplanet. This process
is similar to the gap formation observed in \cite{TanakaIda1997}. Fast migration instead turns
the protoplanet into a predator that plows through the planetesimals
while accreting a small fraction of them. This accretion regime is the focus of
this paper because the case of a single migrating protoplanet can be easily
compared to similar models where core growth is driven by pebble accretion
\citep{Bitsch+etal2015,BitschJohansen2017,Johansen+etal2019}.  A second
motivation for studying single protoplanets rather than oligarchic growth is the
possibility that a few large protoplanets form at ice lines and hence have a
major head start in the accretion process, thus avoiding to compete for 
planetesimals with other protoplanets
\citep{SchoonenbergOrmel2017,DrazkowskaAlibert2017,Liu+etal2019}.

\subsection{Multiple migrating protoplanets}

\cite{ColemanNelson2014} simulated oligarchic growth including migration and
found that the gas giants that form in the simulations had all migrated to the
inner regions of the protoplanetary disc. In other experiments
they showed that gas giants must start their formation beyond 20 AU in the
protoplanetary disc in order to have migration space in which to accrete gas. This
necessity for forming the cores of gas giants very far from the Sun was also
found in pebble accretion models
\citep{Bitsch+etal2015,Ndugu+etal2018,Johansen+etal2019}. \cite{Pirani+etal2019}
proposed that the asymmetric distributions of leading versus trailing Jupiter
trojans is due to the capture of the trojans while Jupiter was migrating rapidly
inwards from its formation region beyond 20 AU.

\subsection{Small planetesimals?}

$N$-body simulations have demonstrated that the formation of cold gas giants is
very hard in all four quadrants of the multiplicity-migration space described
above.  \cite{Rafikov2004} considered the possibility that planetesimals are
either born small or fragment to small sizes. Such small planetesimals (with sizes
smaller than 1 km) have low scale heights (and are hence more easily captured by
the protoplanet) and penetrate planetesimal gaps formed by protoplanet
scattering. The penetration by small planetesimals through the scattering gap
may be the reason why some statistical approaches to the planetesimal population
are able to form a Jupiter-planet on a 5 AU orbit in models neglecting migration and
oligarchic growth \citep{D'Angelo+etal2014}.

Evidence from the solar system nevertheless suggests a characteristic
planetesimal size of 100 km (as we discussed in the introduction). In the
gravitationally unfocused case, the collision timescale of planetesimals with
number density $n_{\rm pla}$, physical cross section $\sigma_{\rm pla}$, radius
$R_{\rm pla}$, internal density $\rho_\bullet$, and relative speed between the
planetesimals $\delta v$ is
\begin{eqnarray}
  t_{\rm coll} &=& \frac{1}{n_{\rm pla} \sigma_{\rm pla} \delta v} \approx
  \frac{R_{\rm pla} \rho_\bullet}{\varSigma_{\rm pla} \varOmega} \nonumber \\
  &=& 50.3\,{\rm Myr}\,f_{\rm pla}^{-1} \left( \frac{R_{\rm pla}}{50\,{\rm km}}
  \right) \left( \frac{\rho_\bullet}{2 \times 10^3\,{\rm kg\,m^{-3}}} \right)
  \left( \frac{r}{10\,{\rm AU}} \right)^{5/2} \, .
\end{eqnarray}
Here we assumed a planetesimal surface density of $\varSigma_{\rm pla} =
10\,{\rm kg\,m^{-2}} f_{\rm pla} [r/(10\,{\rm AU)]^{-1}}$ (corresponding to a
constant 2.4 $M_{\rm E}$ of planetesimals per AU). Even a very massive
planetesimal disc with $f_{\rm pla} = 10$ would have a collision timescale of
1-5 Myr in the 5-10 AU region. Moreover, the characteristic planetesimal size of 100
km observed in the modern solar system indicates that the primordial
planetesimal population did not undergo an extensive fragmentation process to
smaller and
more easily accreted planetesimals.

\subsection{Regions of zero migration?}

By injecting artificial regions of zero-migration in the cold regions of the
protoplanetary disc that trap protoplanets and planetesimals,
\cite{ColemanNelson2016} showed that this approach allows the formation of
cold Jupiters through planetesimal accretion in the outer disc. Their models
successfully produced cold Jupiters because protoplanets and planetesimals
gathered in these convergence zones. However, the mechanism for the emergence of
long-lived zero-migration zones in the formation region of cold gas
giants is not clear, but it might be related to pressure bumps that arise in
non-ideal magnetohydrodynamics (MHD) simulations of protoplanetary discs
\citep{Flock+etal2015,RiolsLesur2019}.

\section{Governing equations}
\label{s:equations}

The main focus of our paper is the simulation of the growth of single
migrating protoplanets by planetesimal accretion. Our approach is similar to
that of \cite{Johansen+etal2019}, who presented core growth tracks for cores
accreting small pebbles alone, but we ignore the contribution of pebbles to core
accretion here. We compare the results obtained here to those of
pebble-driven core accretion models in the conclusions (and in Appendix
A).

\subsection{Planetesimal accretion}

We followed the parametrisations of \cite{TanakaIda1999}, based on analytical fits
to $N$-body simulations, for the planetesimal accretion rate. We briefly summarise these
equations here. The equations are normalised with respect to the local
dynamics near the protoplanet orbiting at a distance $r$ from the star.
Distances are thus normalised by the Hill radius $R_{\rm H}$ of the protoplanet,
while times are normalised by the synodic period $T_{\rm K}/h_{\rm p}$. Here
$T_{\rm K}$ is the Keplerian period and $h_{\rm p} = R_{\rm H} / r = [M/(3
M_\star)]^{1/3}$ is the reduced Hill radius. The physical radius of the
protoplanet is given by $\tilde{R}_{\rm p} = R_{\rm p}/R_{\rm H}$ in these
units. We ignored the contribution of the gas envelope to the cross section of
the protoplanet because the cross section would increase significantly only above a
core mass of 5 $M_{\rm E}$ for our nominal planetesimal radii of 50 km
\citep{D'Angelo+etal2014} and because the accretion rate only scales as the square
root of the effective capture radius for the relevant planetesimal inclinations
considered here (see equation \ref{eq:alphapla}).

The migration timescale is calculated as
\begin{equation}
  \tau_{\rm mig} = \frac{r}{|{\rm d}r/{\rm d} t|} \, ,
\end{equation}
where $r$ is the semi-major axis of the protoplanet and ${\rm d}r/{\rm d}t$ is
the migration speed (see Section \ref{s:planetary_migration} below). The
normalised migration timescale is given by
\begin{equation}
  \tilde{\tau}_{\rm mig} = \tau_{\rm mig} h_{\rm p}^2 / T_{\rm K} \, ,
\end{equation}
where $T_{\rm K}=2\pi/\varOmega$ is the orbital period. This normalisation is a
consequence of the radial position equation $\dot{r} = v_r = r / \tau_{\rm
mig}$.

The gas drag timescale of the planetesimals is
\begin{equation}
  \tau_{\rm gas} = \frac{2 M_{\rm pla}}{C_{\rm D} \pi R_{\rm pla}^2 \rho_{\rm g}
  v_{\rm K}} \, ,
\end{equation}
where $M_{\rm pla}$ is the planetesimal mass, $C_{\rm D}$ is the drag
coefficient, $\rho_{\rm g}$ is the gas density, and $v_{\rm K}$ is the Keplerian
speed of the protoplanet. We set $C_{\rm D}=0.5$ for simplicity; this non-linear
drag regime is valid for planetesimals larger than a few kilometers
\citep{Weidenschilling1977}. The normalised gas-drag timescale, which enters
the damping equations for eccentricity and inclination, is then given by
\begin{equation}
  \tilde{\tau}_{\rm gas} = \tau_{\rm gas} / T_{\rm K} \, .
\end{equation}
\cite{TanakaIda1999} found a critical migration timescale for shepherding,
\begin{equation}
  \tilde{\tau}_{\rm mig,c} = 0.81 \left( \sqrt{1+0.45(\tilde{\tau}_{\rm
  gas})^{2/3}} + 1 \right)^2
.\end{equation}
Protoplanets with $\tilde{\tau}_{\rm mig} > \tilde{\tau}_{\rm mig,c}$ are
considered to be shepherds and stop their accretion of planetesimals. These
shepherding bodies nevertheless continue to migrate into regions of higher gas
density, which in turn increases the gas drag on the planetesimals and hence
reinforces the shepherding behaviour (the {\it \textup{run-away shepherding}} effect).

The accretion rate of the protoplanet is given by
\begin{equation}
  \dot{M} = \mathcal{E} \dot{M}_{\rm pla} = \mathcal{E} 2 \pi r \dot{r}
  \varSigma_{\rm pla} \, .
\end{equation}
Here $\mathcal{E}$ is the accretion efficiency and $\dot{M}_{\rm pla}$ is the
flux of planetesimals that cross the orbit of the migrating protoplanet.
The migration speed is parameterised through the inverse of the normalised
migration timescale,
\begin{equation}
  \dot{\tilde{b}}_{\rm p} = \tilde{\tau}_{\rm mig}^{-1} \, .
\end{equation}
The accretion efficiency is then parameterised as
\begin{equation}
  \mathcal{E} = \alpha_{\rm pla} \dot{\tilde{b}}_{\rm p}^{\beta_{\rm pla}-1} \,
  .
  \label{eq:eff}
\end{equation}
The parameters $\alpha_{\rm pla}$ and $\beta_{\rm pla}$ are fits to numerical
simulations.  \cite{TanakaIda1999} found the expressions
\begin{eqnarray}
  \alpha_{\rm pla} &=& 2.5 \sqrt{\frac{\tilde{R}_{\rm p}}{1+0.37
  \tilde{i}_0^2/\tilde{R}_{\rm p}}} \, ,
  \label{eq:alphapla} \\
  \beta_{\rm pla} &=& 0.79 \left( 1 + 10 \tilde{i}_0^2 \right)^{-0.17} \, .
\end{eqnarray}
Here $\tilde{i}_0$ is the inclination of the planetesimal population, divided by
$h_{\rm p}$ to convert this into Hill sphere units.

\subsection{Planetesimal inclinations}

\cite{TanakaIda1999} showed that a protoplanet easily excites the eccentricities
of the planetesimals that it migrates towards, but the inclination of a
planetesimal is hardly changed prior to the first scattering by the protoplanet.
Therefore the planetesimal inclination is an important parameter in the
accretion rate. We mainly set the inclination from the balance between the
stirring by the turbulent gas and damping by gas drag, but for our experiments
with a very or extremely weak turbulent stirring, we also included the inclination
excitation by mutual planetesimal scatterings (see below).

\cite{Ida+etal2008} used a non-dimensional parameter $\gamma$ to quantify the
stirring by the turbulent gas and find the equilibrium eccentricity through
\begin{eqnarray}
  e_{\rm drag} &=& 0.013 f_{\rm g}^{1/3} \left( \frac{\gamma}{10^{-4}}
  \right)^{2/3} \left( \frac{R}{50\,{\rm km}} \right)^{1/3} \nonumber \\
  && \qquad \qquad \times \left( \frac{\rho_\bullet}{2 \times 10^3\,{\rm
  kg\,m^{-3}}}\right)^{1/3} \left( \frac{r}{10\,{\rm AU}} \right)^{11/12} \, .
  \label{eq:edrag}
\end{eqnarray}
Here $f_{\rm g}$ is the gas surface density profile relative to the minimum mass
solar nebula \citep[$\varSigma_{\rm g}=2.4\times10^{4}\,{\rm kg\,m^{-2}} (r/{\rm
AU})^{-1.5}$ is used as a reference value by][]{Ida+etal2008}. We set $i_0 =
e_{\rm drag}$ because turbulent density fluctuations appear to excite
eccentricities and inclinations equally well \citep{Yang+etal2012}.

The turbulent stirring parameter $\gamma$ that appears in equation
(\ref{eq:edrag}) gives the scaling factor of the random-walk excitation of the
planetesimal eccentricity. \cite{Ida+etal2008} suggested that $\gamma \sim
\alpha_{\rm t}$ (where $\alpha_{\rm t}$ here denotes the turbulent diffusion
coefficient, in contrast to the $\alpha$ value that controls the viscous
evolution of the protoplanetary disc), or $\gamma \sim \alpha_{\rm t}^{1/2} H/r$
(where $H/r$ is the disc aspect ratio). Observations of the thickness of the
dust layer in the protoplanetary disc around the young star HL Tau suggest a
turbulent diffusion coefficient of a few times $10^{-4}$ \citep{Pinte+etal2016},
but the interpretation depends on the assumed Stokes number of the sedimented
pebbles. The radial scale height of pebble rings observed in the  Disk
Substructures at High Angular Resolution Project (DSHARP) indicates a diffusion
coefficient $\sim$$10^{-3}$ \citep{Dullemond+etal2018}. The physical origin of
this weak turbulence is unknown; the vertical shear instability and active
surface layer turbulence both yield turbulent diffusion coefficients in excess
of $10^{-3}$ in the mid-plane \citep{StollKley2016,Yang+etal2018}. We here
assumed a relatively low value of $\gamma = 10^{-4}$ as the nominal value
because this is consistent with observations and likely represents a lower limit
to the realistic span of values. This value of $\gamma$ also yields planetesimal
inclinations similar those measured in simulations of the stirring of
planetesimals in the dead zone by density waves propagating down from turbulent
surface layers \citep{Gressel+etal2012}.  However, in Section \ref{s:boost} we
show the change in our results when we adopt even lower values of the stirring
coefficient.

\cite{Ida+etal2008} also argued that the highest eccentricity induced by mutual
planetesimal scatterings is reached when the random velocity becomes comparable
to the escape speed because this situation favours collisions over strong
scattering. The resulting eccentricity only depends on the planetesimal escape
speed and on the Keplerian speed,
\begin{eqnarray}
  e_{\rm acc} &=& \frac{v_{\rm esc}}{v_{\rm K}} = 0.0046\,\left(
  \frac{R}{50\,{\rm km}} \right) \nonumber \\
  & & \qquad \qquad \times \left( \frac{\rho_\bullet}{2 \times 10^3\,{\rm
  kg\,m^{-3}}}\right)^{1/2} \left( \frac{r}{10\,{\rm AU}} \right)^{1/2} \, .
  \label{eq:eacc}
\end{eqnarray}
Our nominal value of $\gamma=10^{-4}$ in equation (\ref{eq:edrag})
therefore represents the lower end of the range where mutual scatterings are less
important than turbulent stirring, for a planetesimal radius of 50 km. We
therefore ignored for simplicity inclination excitation by planetesimal
scatterings for the main simulations in Section \ref{s:growth}, but we included
it for consistency in our experiments with lower values of $\gamma$ presented in
Section \ref{s:boost}.

\subsection{Planetary migration}
\label{s:planetary_migration}

We used a combined expression for planetary migration that captured the nominal
type I migration of low-mass cores and the transition to migration modified by
gap formation \citep[based on][]{Kanagawa+etal2018}. This migration prescription
and its consequences for planet formation are discussed in \cite{Ida+etal2018}
and in \cite{Johansen+etal2019}. The migration is proportional to the gas
surface density and initially to the planetary mass, but the migration rate
drops as the inverse of the planetary mass after the protoplanet reaches a
gap-opening mass of approximately $20$ $M_{\rm E}$. The turbulent viscosity,
which determines the gap-opening mass, was set to a low value of $\alpha_{\rm
t}=10^{-4}$.

\subsection{Gas accretion}

\cite{Johansen+etal2019} allowed gas accretion to commence after the core
reached the pebble isolation mass; here we instead assumed that a core mass of 10
$M_{\rm E}$ marks the transition to gas accretion. This simplified approach
guarantees a core mass that is comparable to those inferred from the giant
planets in the solar system \citep{Guillot2005,Wahl+etal2017}, while it allows
us emphasise the core formation process rather than the physics of
gas accretion. The expressions that we use for the gas accretion rate are
discussed in detail in \cite{Ida+etal2018} and \cite{Johansen+etal2019}. We
chose a low opacity in the gas envelope to facilitate rapid gas accretion; the
gas accretion timescale of a Jupiter-mass planet is generally a few hundred
thousand years in this model.

\subsection{Evolution of the protoplanetary disc}

The protoplanetary disc is evolved as a standard viscous $\alpha$ disc with
$\alpha=0.01$. The disc surface density profile follows a (negative) power law
out to a characteristic radius beyond which the profile is exponentially
tapered.  We used analytical expressions for the evolution of the mass accretion
rate onto the star and of the characteristic disc radius, anchored by choosing
the initial gas mass flux onto the star, $\dot{M}_0$, and the final gas mass
flux after 3 Myr of disc evolution, $\dot{M}_3$, after which the protoplanetary
disc is assumed to photoevaporate. The choice of the final accretion rate
effectively sets the initial characteristic disc size; a high value of
$\dot{M}_3$ implies a large initial disc size and vice versa
\citep[see][]{Johansen+etal2019}.

We ignored viscous heating of the protoplanetary disc, motivated by simulations
showing that this heating is weak in discs accreting by magnetic stresses
\citep{Mori+etal2019}. We therefore considered only stellar irradiation to
set the fixed gas temperature profile, using the irradiated temperature
profile given in \cite{Ida+etal2016}.  This temperature approach yields a disc
aspect ratio $H/r=0.024$ at 1 AU that increases with distance as a power law of
index $2/7$.  The smooth disc profile precludes any regions of outwards
migration in the disc, but such regions are limited to the inner few AU
from the star in any case \citep[unless the turbulent viscosity is high,
see][]{Cossou+etal2014,BitschJohansen2016} and hence do not affect the growth
tracks for cold gas giants, which are the focus of this paper.
\begin{figure*}
  \begin{center}
    \includegraphics[width=0.9\linewidth]{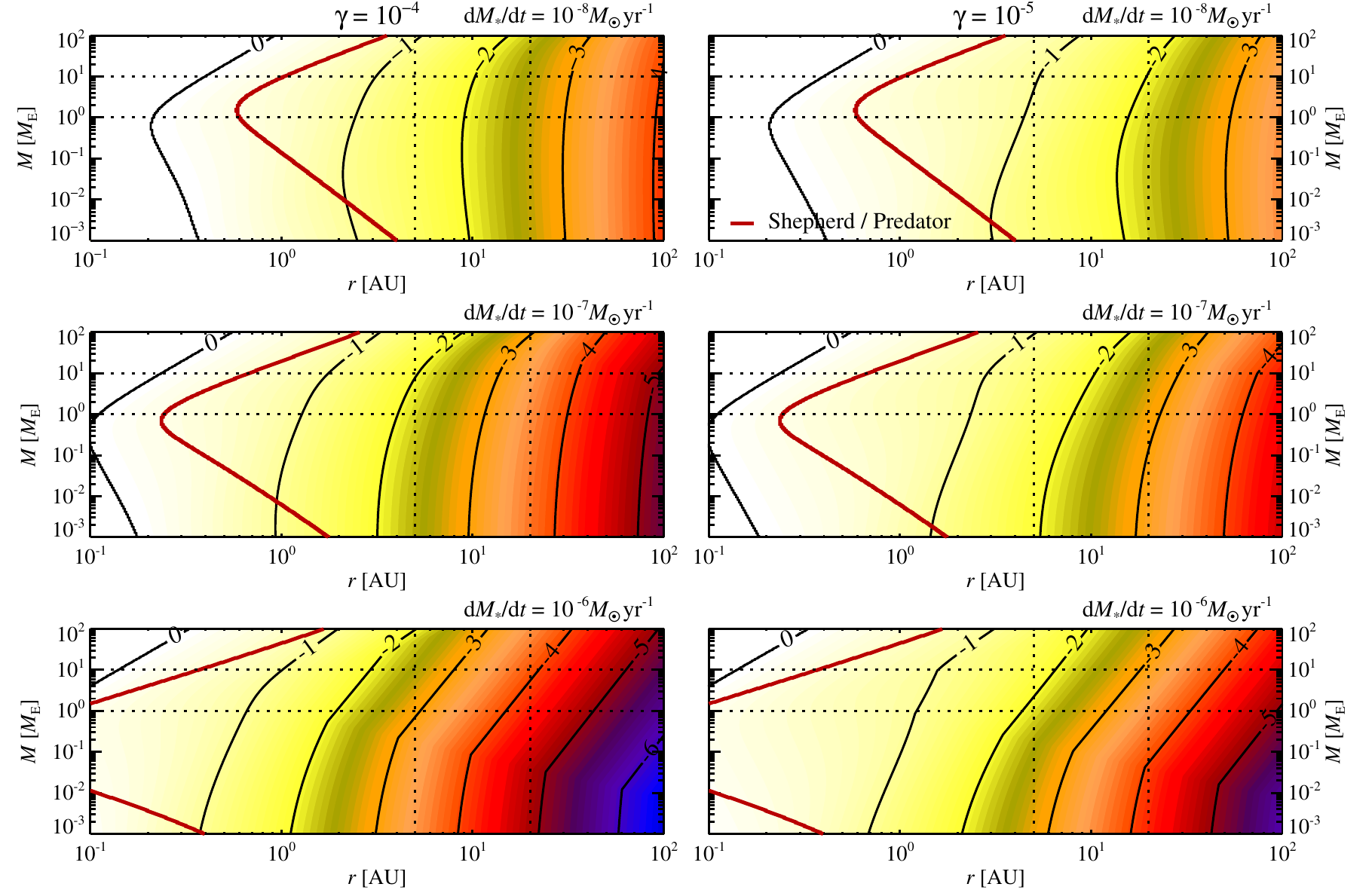}
  \end{center}
  \caption{Accretion efficiency of a protoplanet accreting planetesimals of
  radius 50 km and for weak turbulent stirring ($\gamma=10^{-4}$, left plots)
  and very weak turbulent stirring ($\gamma=10^{-5}$, right plots) as a
  function of the distance from the star and the protoplanet mass. The three
  rows show the accretion efficiency for three different increasing (top to
  bottom) values of the mass accretion rate through the disc; the gas surface
  density in turn defines the equilibrium inclination of the planetesimals and
  the transition of the protoplanet from an accreting predator (exterior of the
  red line) to a non-accreting shepherd (interior of the red line). The
  parameter region of core formation in the solar system (the box made by
  5 and 20 AU and by 1 and 10 $M_{\rm E}$  ) is marked with dashed lines. The
  accretion efficiency beyond 5 AU lies at a value of a few percent or lower
  for the case of weak turbulent stirring, while lowering the turbulent stirring
  to $\gamma=10^{-5}$ increases the efficiency by approximately a factor 3.  The
  efficiency overall increases with decreasing gas accretion through the disc
  because more slowly migrating planets accrete a higher fraction of the planetesimals
  that they migrate through (even though the planetesimal inclinations 
  increase slightly as the gas drag is reduced).}
  \label{f:filtering_fraction}
\end{figure*}

\section{Growth of single migrating protoplanets}
\label{s:growth}

\subsection{Accretion efficiency}

We start by calculating the efficiency of accreting planetesimals of 50 km in
radius (100 km in diameter) using the \cite{TanakaIda1999} parametrisation for
two different values of the turbulent stirring (weak stirring with
$\gamma=10^{-4}$ and very weak stirring with $\gamma=10^{-5}$) as a function of
the distance from the star and the mass of the protoplanet in Figure
\ref{f:filtering_fraction}. The planetesimal inclinations were calculated taking
only the turbulent stirring by the gas into account. The results are shown for
three values of the stellar gas accretion rate, $\dot{M}=10^{-6}\,M_\odot\,{\rm
yr}^{-1}$, $\dot{M}=10^{-7}\,M_\odot\,{\rm yr}^{-1}$ , and
$\dot{M}=10^{-8}\,M_\odot\,{\rm yr}^{-1}$. We considered here a very large disc
size, such that $\dot{M}$ is inwards and constant out to 100 AU. We recall that we ignored viscous heating; the inclusion of viscous heating would
heat the inner regions of the protoplanetary disc and hence decrease the
migration rate, which would in turn make the protoplanets more prone to
shepherding. The considered accretion rates represent an evolutionary sequence
in which the accretion rate decreases with increasing age of the protoplanetary
disc. The stellar accretion rate affects the planetesimal accretion efficiency
mainly through the gas surface density, which in turn scales the migration rate
and the inclination of the planetesimals in equilibrium between turbulent
stirring and gas damping.

The accretion efficiency for planetesimals of $50$ km in radius lies in the
range between $0.0001$ and $0.03$ in the region of giant planet formation in the
solar system (the region between 5 and 20 AU in distance from the star and 1 and
10 $M_{\rm E}$ in mass is marked with dotted lines in the plots). Lowering the
turbulent stirring coefficient to $\gamma=10^{-5}$ increases the efficiency by
approximately a factor three. This increase comes mainly from $\alpha_{\rm
pla} \propto \tilde{i}_0^{-1}$ (equation \ref{eq:alphapla} for $\tilde{i}_0 \gg
\tilde{R}_{\rm p}$) and $\tilde{i}_0 \propto \gamma^{2/3}$ (equation
\ref{eq:edrag}), which combine to make $\mathcal{E} \propto \gamma^{-2/3}$
(equation \ref{eq:eff}). This scaling slightly overpredicts the actual increase
when lowering $\gamma$ from $10^{-4}$ to $10^{-5}$ because of saturation
effects in equation (\ref{eq:alphapla}) towards the 2D limit. The efficiency
increases with decreasing stellar mass accretion rate because more slowly migrating
protoplanets accrete a larger fraction of the planetesimal flux.
\begin{figure*}
  \begin{center}
    \includegraphics[width=0.9\linewidth]{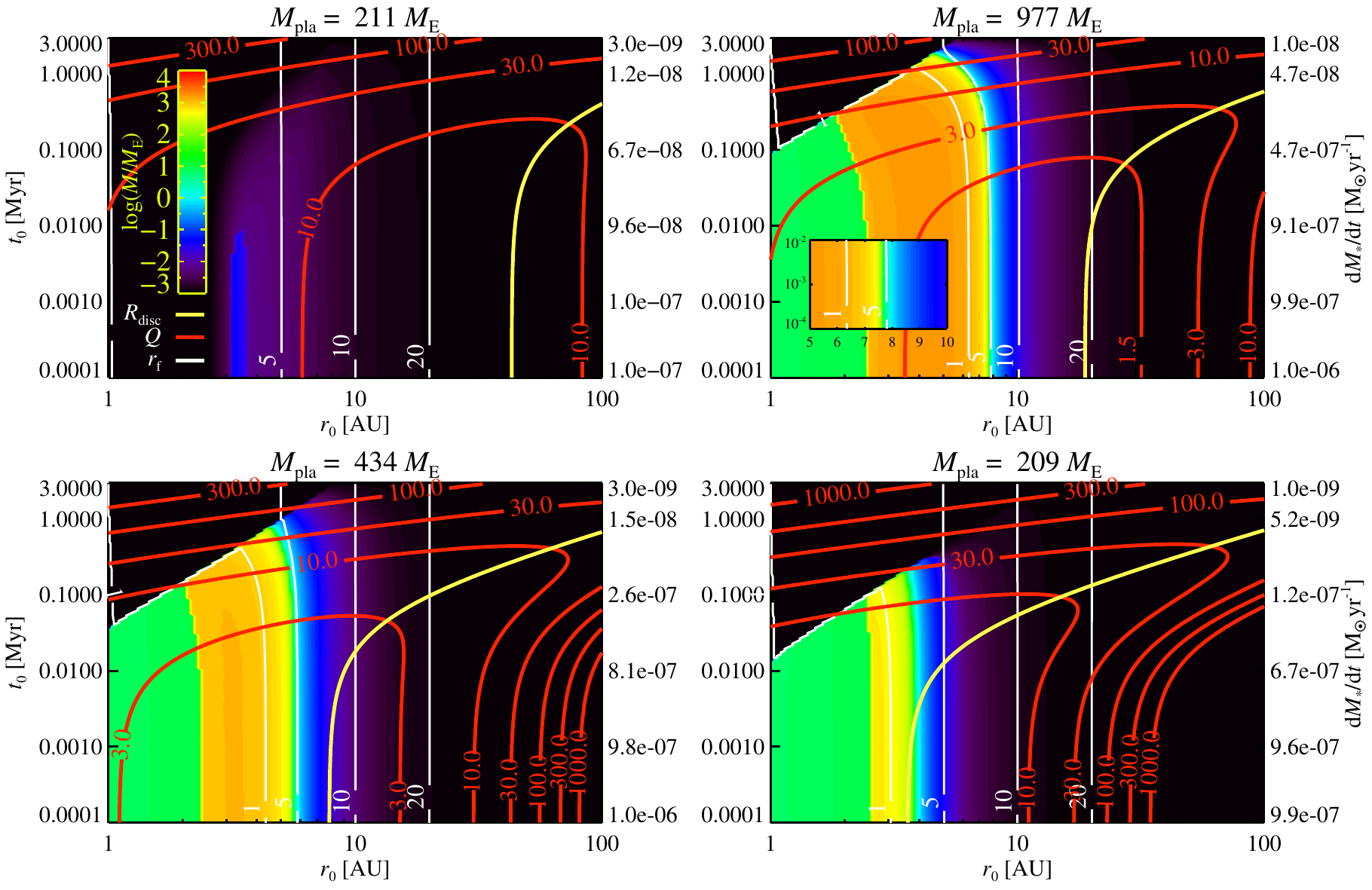}
  \end{center}
  \caption{Growth maps for protoplanet growth by planetesimal accretion, with
  gas accretion commencing after 10 $M_{\rm E}$ is reached. The maps show the
  final mass (i.e. the mass reached when the disc has dissipated after 3 Myr)
  as continuous colours and the final positions as white contour lines as a
  function of the starting position of the protoplanet and the starting time. We
additionally show the size of the protoplanetary disc (yellow line) and the
  Toomre $Q$ of the gas (red contour lines). The panels show the results for
  four values of the initial mass accretion rate through the disc and the mass
  accretion rate after 3 Myr. The instantaneous stellar mass accretion rate is
  shown on the right axis. Each panel is also labelled with the total mass of the
  planetesimals in the models; we assume here that 1\% of the initial gas
  surface density profile is converted into planetesimals at $t=0$.  The nominal
  model (upper left) achieves protoplanet growth only up to 0.1 $M_{\rm E}$,
  approximately the mass of Mars. The three other models, where planetesimals
  form at the gravitationally unstable phase of protoplanetary disc evolution,
  successfully form hot and warm gas giants interior of 2 AU, but are not able
  to form any cold gas giants in the 5--10 AU region (see inset in the upper
  right panel).} \label{f:growth_map_planetesimals}
\end{figure*}
\begin{figure*}
  \begin{center}
    \includegraphics[width=0.9\linewidth]{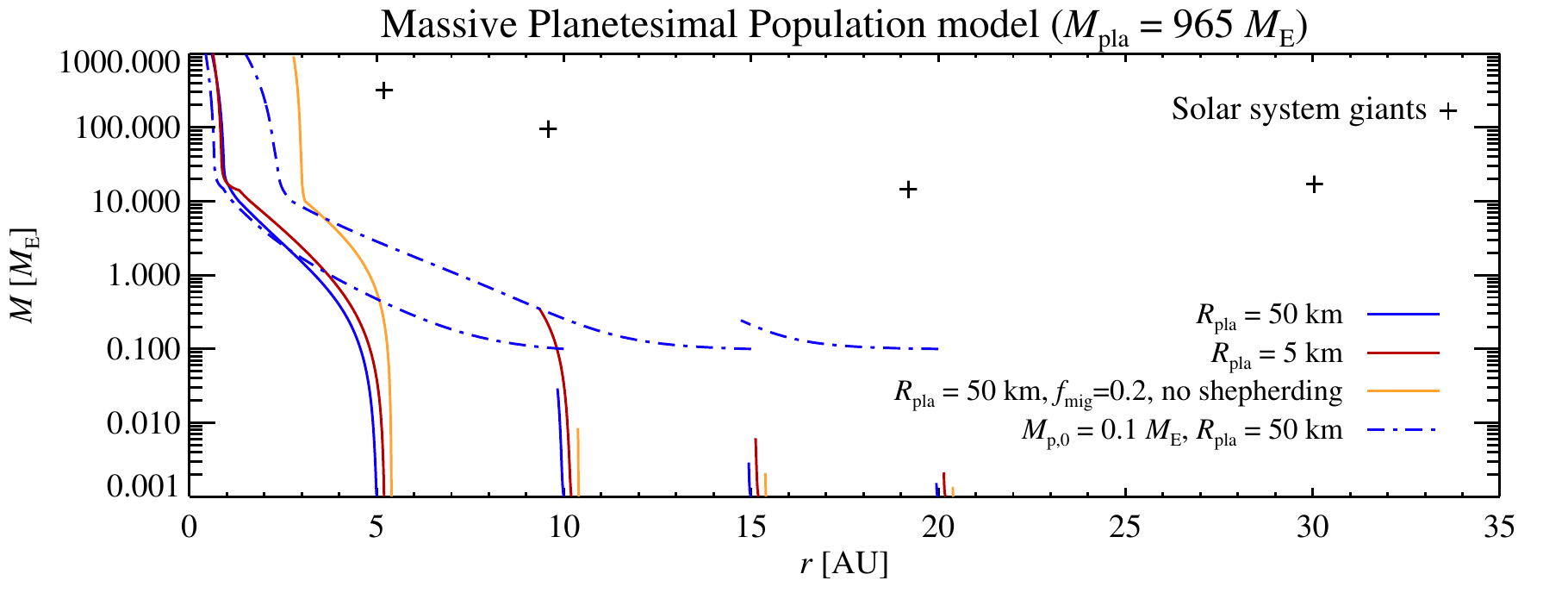}
  \end{center}
  \caption{Growth tracks of protoplanets accreting planetesimals of two
  different radii (blue and red lines), with reduced migration (yellow lines)
  and starting with a higher protoplanet mass (blue dash-dotted lines). The
  growth tracks are started at either 5, 10, 15, or 20 AU. We used here the
  massive planetesimal population model that contains 965 $M_{\rm E}$ of
  planetesimals formed at $t=0$ and assumed that the protoplanets start to grow
  immediately thereafter. Growth tracks starting at 5 AU generally become
  warm gas giants, ending in final orbits around 1 AU.  Protoplanets starting
  farther out experience only moderate growth due to the low planetesimal
  accretion efficiencies there. Lowering the migration rate has a net positive
  effect for the growth track starting at 5 AU, but the more distant
  protoplanets experience a decreased growth rate when they migrate more
  slowly because the flux of planetesimals past migrating the protoplanet is
  reduced. The more massive protoplanets experience rapid migration towards the
  regions of high accretion efficiency close to the star, but the adoption
  of initially more massive protoplanets does not help in forming cold gas
  giants.}
  \label{f:growth_tracks_planetesimals_Rpla}
\end{figure*}

The transition from predator to shepherd is marked with a thick red line in
Figure \ref{f:filtering_fraction}. This transition depends on the gas drag
timescale and on the migration speed. The transition to shepherding
generally occurs at around 1 AU from the star in the final decade of core growth ($M
\approx 1-10$ $M_{\rm E}$). Cores with lower mass transition to shepherding farther
out.

\subsection{Growth map}

We calculated a large number of growth tracks, defined by the starting position
$r_0$ and starting time $t_0$ of the protoplanet. We started the protoplanets at
a mass of $10^{-3}$ $M_{\rm E}$, a value that may even be reached by the most
massive planetesimals formed by the streaming instability \citep{Liu+etal2019}.
We considered four different combinations of the beginning and final stellar
accretion rate: $(\dot{M}_0,\dot{M}_3)=$ $(10^{-7},3\times10^{-9})$,
$(10^{-6},10^{-8})$, $(10^{-6},3\times10^{-9})$, and $(10^{-6},10^{-9})$. The
planetesimals were assumed to form at $t=0$ following the primordial gas surface
density profile. With a planetesimal metallicity of $Z_{\rm pla}=0.01$, these
protoplanetary disc models contain $187$ $M_{\rm E}$, $965$ $M_{\rm E}$, $432$
$M_{\rm E}$ , and $208$ $M_{\rm E}$ of planetesimals, respectively. The
planetesimals were assumed to be 50 km in radius and the turbulent stirring
coefficient was chosen at its nominal value of $\gamma=10^{-4}$ for all four
growth maps.

The growth maps in Figure \ref{f:growth_map_planetesimals} indicate the final
mass of the planets in continuous colours and the final positions as white
contour lines. The final mass decreases for protoplanets starting at later times
because the time left for growing the planet is shorter. Protoplanets starting their
growth close to the star or late in the disc evolution migrate slowly enough to
become planetesimal shepherds; hence the final mass drops to zero at late
starting times. The yellow line marks the size of the protoplanetary disc and
the red contours show the self-gravity parameter $Q=c_{\rm s} \varOmega/(\pi G
\varSigma_{\rm g})$ \citep{Safronov1960,Toomre1964}, where $c_{\rm s}$ is the
sound speed of the gas, $\varOmega$ is the Keplerian frequency, and
$\varSigma_{\rm g}$ is the gas surface density.

The protoplanetary disc model shown in the upper left corner contains
approximately twice the mass in planetesimals of the minimum mass solar nebula
\citep{Hayashi1981}, consistent with the constraints on the primordial
planetesimal populations discussed in the introduction. This model 
entirely fails to form any gas giants. Because the planetesimal accretion efficiency
is at most a few percent, the total planetesimal mass of $\approx$187 $M_{\rm
E}$ is insufficient to form a core of 10 $M_{\rm E}$. The most massive
protoplanets grow to approximately $0.1$ $M_{\rm E}$ (comparable to Mars) in
this model.

When the initial mass accretion rate is increased by a factor 10 to $10^{-6}$
$M_\odot\,{\rm yr}^{-1}$, the model shown in the upper right corner of Figure
\ref{f:growth_map_planetesimals} yields a protoplanetary disc that starts with a
self-gravity parameter $Q$ below 1.5 in a large region from 3 to 30 AU. The disc
initially has a characteristic size of 20 AU, beyond which the power-law gas
surface density transitions to an exponential tapering. This model successfully
produces gas-giant planets that end in orbits up to 2 AU from the star, but
fails to form any analogues of Jupiter and Saturn in the 5--10 AU region.

The two lower plots of Figure \ref{f:growth_map_planetesimals} show versions of
the protoplanetary disc with an initially high accretion rate, but with smaller
initial disc sizes (9 AU and 4 AU) and lower final accretion rates
($\dot{M}_3=3\times10^{-9}\,M_\odot\,{\rm yr}^{-1}$ and
$\dot{M}_3=10^{-9}\,M_\odot\,{\rm yr}^{-1}$). The reduction in
initial disc size pushes the final positions of the gas giants that are formed
even closer to the star. Protoplanets can grow even outside of the
initial disc size because we assumed that planetesimals also form at an abundance
of 1\% relative to the gas in the exponentially tapered regions.
\begin{figure}
  \begin{center}
    \includegraphics[width=0.8\linewidth]{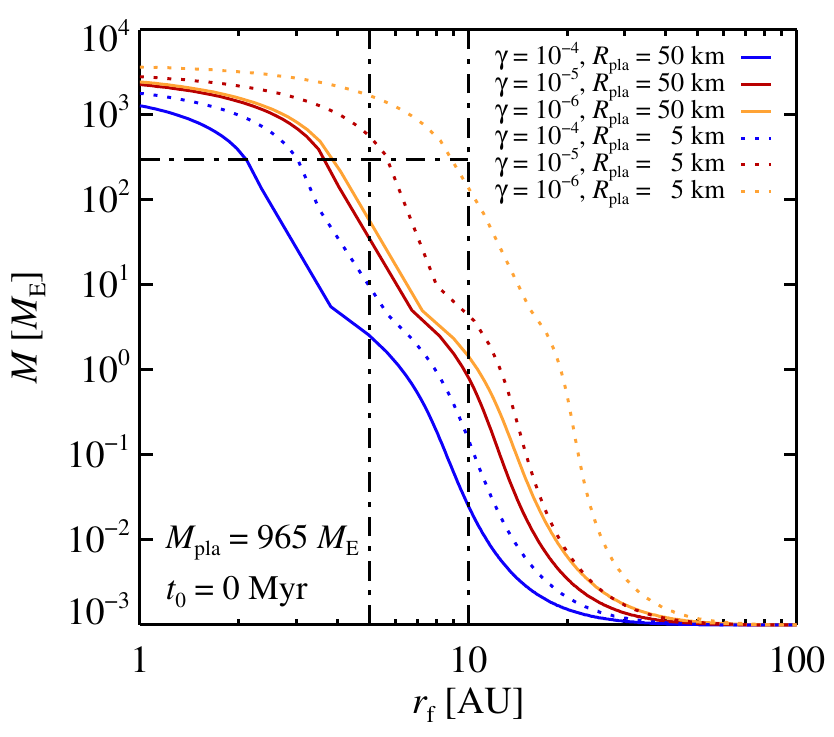}
  \end{center}
  \caption{Final planetary mass as a function of the final planetary orbit
  for protoplanets starting to accrete at $t=0$ in the massive planetesimal
  population model with two different planetesimal radii (50 km and 5 km) and
  three values of the turbulent stirring coefficient ($10^{-4}$, $10^{-5}$, and
  $10^{-6}$). The solid blue line ($\gamma=10^{-4}$ / $R_{\rm pla}=50\,{\rm
  km}$) corresponds to a cut at $t_0=0$ of the upper right panel of Figure
  \ref{f:growth_map_planetesimals}. The planetesimal inclinations are here set
  as the maximum of the inclinations given by turbulent stirring and by mutual
  planetesimal scattering. Planetesimals with a nominal radius of 50 km can only
  form Jupiter-mass planets interior of 3 AU, even when the turbulent stirring
  is very weak with $\gamma=10^{-5}$ or extremely weak with $\gamma=10^{-6}$,
  because of inclination stirring by mutual planetesimal scattering. Lowering
  the planetesimal radius to 5 km allows the formation of Jupiter analogues when
  the turbulent stirring coefficient is $\gamma=10^{-5}$ or $\gamma=10^{-6}$.}
  \label{f:MMtf_rrf_tgamma_Rpla}
\end{figure}

\subsection{Boosting planetesimal accretion}
\label{s:boost}

We tested ways to boost core growth by planetesimal accretion by decreasing
the size of the planetesimals, decreasing the migration speed, increasing the
initial mass of the protoplanets, and decreasing the inclinations of the
planetesimals.
We illustrate the growth tracks of protoplanets starting at 5, 10, 15, and 20 AU
in Figure \ref{f:growth_tracks_planetesimals_Rpla} in what we call the massive
planetesimal population model (the model at the top right of Figure
\ref{f:growth_map_planetesimals}: initial size of 20 AU, initial stellar
accretion rate of $10^{-6}$ $M_\odot\,{\rm yr}^{-1}$ , and 965 $M_{\rm E}$ of
planetesimals) for two different planetesimal sizes (50 km and 5 km), for a case
where the migration speed is reduced, and for a case with a more massive initial
protoplanet. We also overplot the current positions of the giant planets in the
solar system.  Starting at 5 AU and with planetesimals of 50 km, we can only form
Jupiter-mass planets ending in (warm) 1 AU orbits. Starting farther out at 10
AU, smaller planetesimals help in promoting the growth of the protoplanet, but the nominal planetesimal radius of 50 km and the smaller planetesimals with
radius 5 km both fail to produce protoplanets beyond $1$ $M_{\rm E}$ in mass.

In order to explore the role of the migration rate on the outcome of planet
formation, we produced additional growth tracks with the migration multiplied by
a reduction factor $f_{\rm mig}=0.2$ for all protoplanet masses. We also ignored
the transition to shepherding in this experiment because shepherding might be
suppressed when multiple protoplanets scatter planetesimals between them
\citep{KokuboIda2000,Thommes+etal2003}. The result is shown with yellow curves
in Figure \ref{f:growth_tracks_planetesimals_Rpla}. The reduced migration has a
weak positive effect on the growth tracks starting at 5 AU; they now produce
gas giants in 2--3 AU orbits. The effect is overall negative farther out, however.
Lowering the migration speed leads to a higher efficiency of accreting
planetesimals, but the flux of planetesimals that the protoplanets migrates over
is also reduced, which reduces the overall growth rate.

In Figure \ref{f:growth_tracks_planetesimals_Rpla} we show with dash-dotted blue
lines the growth tracks when we started with a larger protoplanet mass of $M_{\rm
p,0}=0.1\,M_{\rm E}$. These protoplanets experience rapid migration towards the
star. This has the positive effect of bringing the protoplanets into the regions of
higher accretion efficiency closer to the star, but increasing the protoplanet
mass does not help to create giant planets in cold orbits.

We show the final planet mass as a function of the final position forming in the
massive planetesimal population model in Figure \ref{f:MMtf_rrf_tgamma_Rpla}
for lower values of the turbulent stirring $\gamma$. We assumed again that the
protoplanets start growing immediately at $t=0$. We include here the inclination
stirring by mutual scattering (equation \ref{eq:eacc}) and by torques from
the turbulent gas (equation \ref{eq:edrag}, multiplied by 0.5 because $i \approx
0.5 e$ is typical for excitation by scattering). The planetesimal inclination is
set as the larger of the two expressions. We tested three values of the turbulent
stirring coefficient $\gamma$ (weak turbulence: $\gamma=10^{-4}$, very weak
turbulence: $\gamma=10^{-5}$, and extremely weak turbulence: $\gamma=10^{-6}$) and
two planetesimal sizes (50 km and 5 km). The outermost Jupiter-mass planet
shifts from 2 AU to 3 AU when the turbulence strength is decreased from weak to very
weak or extremely weak. In the latter two cases the inclination is mainly set by
the escape speed of the planetesimals. Lowering the planetesimal radius to 5 km
permits the formation of Jupiter-mass planets in the 5 AU region when the
turbulence is very weak or extremely weak.

\subsection{Later planetesimal formation}

\begin{figure}
  \begin{center}
    \includegraphics[width=0.9\linewidth]{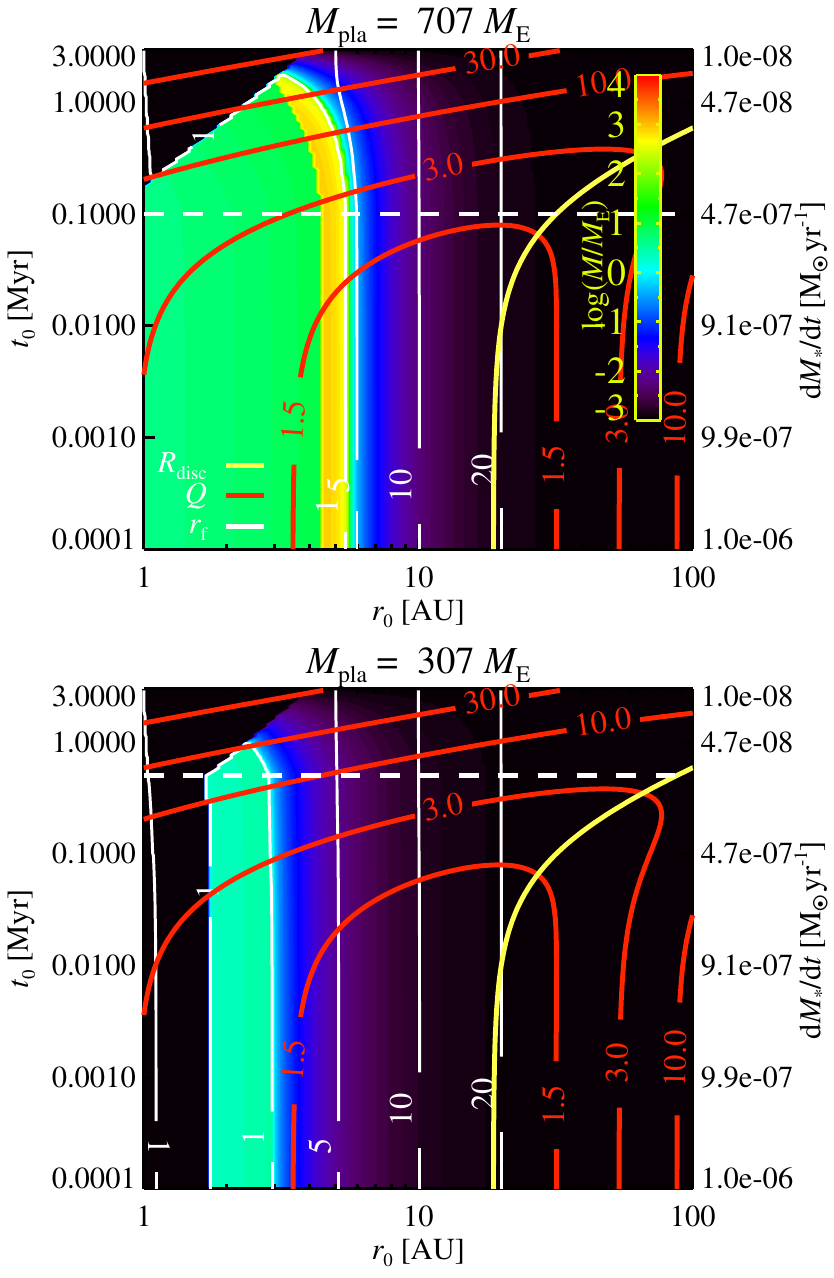}
  \end{center}
  \caption{Growth maps for core growth by accretion of 50 km
  planetesimals that form at either $t=0.1$ Myr (top panel) or $t=0.5$ Myr
  (bottom panel). The planetesimal population is given 1\% of the gas surface
  density at the evolution stage of the protoplanetary disc at the chosen
  formation time. The protoplanets are nevertheless still started at the full
  range of starting times, even before the planetesimals form. Delaying the
  formation of planetesimals to $t=0.1$ Myr reduces the parameter space for
  forming hot and warm gas giants significantly, while the delay to $0.5$ Myr
  entirely removes the capability of the protoplanetary disc to form any giant
  planets.}
  \label{f:growth_map_planetesimals_tpla}
\end{figure}

Iron meteorites are evidence that some planetesimal populations formed very
early in the solar system and subsequently melted by heat from the decay of
short-lived radioactive elements \citep{Bizzarro+etal2005,Kleine+etal2009}.
However, most meteorites are classified as primitive and undifferentiated and
hence likely formed at least 1 Myr after the formation of the Sun, so that
melting and differentiation were avoided \citep{Larsen+etal2016}. We therefore explore in
Figure \ref{f:growth_map_planetesimals_tpla} the effect of forming planetesimals
at either $t=0.1$ Myr or at $t=0.5$ Myr. Delaying the formation of planetesimals
by 0.1 Myr has no significant effect on the growth map because the protoplanetary
disc is still in its pristine phase at this time and forms a massive population
of 707 $M_{\rm E}$ of planetesimals.  On the other hand, delaying planetesimal
formation to 0.5 Myr and forming 307 $M_{\rm E}$ of planetesimals is
detrimental to the growth of the protoplanet and entirely removes the
possibility of forming gas-accreting cores.

\section{Summary and discussion}
\label{s:summary}

We have used the fits to $N$-body simulations of single migrating
protoplanets from \cite{TanakaIda1999} to produce growth tracks for protoplanets
growing by accreting planetesimals. We assumed that gas accretion is rapid and
commences when the core reaches a mass of 10 Earth masses. We evolved the
protoplanetary gas disc using the standard viscous approach. The initial disc
size was set by defining the beginning stellar accretion rate and the final
stellar accretion rate after three million years when we assume that the remaining
gas is photoevaporated.

Our nominal parameters for the planetesimal population followed from three
constraints that come from studies of the solar system and protoplanetary discs.
The first constraint comes from theoretical models of terrestrial planet
formation and from the inferred modern populations of planetesimals in Neptune's
scattered disc and in the Oort cloud; these models both indicate a primordial
planetesimal population in the solar system of a few Earth masses per
AU \citep{Raymond+etal2009,Brasser2008,BrasserMorbidelli2013}.
The second constraint is that the characteristic planetesimal diameter in the
solar system, inferred from the asteroid belt and from the Kuiper belt
\citep{Bottke+etal2005,Morbidelli+etal2009,Abod+etal2018,Singer+etal2019,Stern+etal2019},
is approximately 100 km; we take this as the nominal birth size of the
planetesimals. The stirring of the planetesimal inclinations by the turbulent
gas plays a very important role in setting the efficiency of planetesimal
accretion. Our third constraint comes from protoplanetary discs; here the
turbulent stirring coefficient of dust is inferred to lie in the range between
$10^{-4}$ and $10^{-3}$ \citep{Pinte+etal2016,Dullemond+etal2018}. We therefore
used a nominal value $\gamma=10^{-4}$ for the turbulent stirring coefficient;
this value is also consistent with measurements of the planetesimal inclinations
in the dead zone of protoplanetary discs \citep{Gressel+etal2012}.

Based on these constraints, we built a nominal protoplanetary disc model where
planetesimals formed when the stellar accretion rate is $10^{-7}\,M_\odot\,{\rm
yr}^{-1}$ and the disc has an initial size of 40 AU. The disc
contains 187 Earth masses of planetesimals, slightly higher than twice the
minimum mass solar nebula \citep{Hayashi1981}. However, the protoplanets in this
model grow only to a maximum mass of 0.1 Earth masses, similar to the mass of
Mars. The same protoplanetary disc model yielded super-Earths and gas-giant
planets when the core mass was evolved only through the accretion of a similar
amount of small pebbles \citep{Johansen+etal2019}.

We therefore tested another model where the planetesimals formed already at the
self-gravitating stage of protoplanetary disc evolution, where the mass
accretion onto the star was $\dot{M}=10^{-6}\,M_\odot\,{\rm yr}^{-1}$. This instead yielded 965 Earth masses of planetesimals in a protoplanetary disc of an
initial size of 20 AU; this enormous mass, corresponding to 50
Earth masses per AU or approximately 13 times the minimum mass
solar nebula, is in clear conflict with the inferred primordial population of
planetesimals in the solar system (the first constraint above). This model
successfully produced hot and cold gas giants, but failed to form any analogues
of Jupiter and Saturn in orbit and mass. We experimented with smaller
planetesimals and lower values of the turbulent stirring and were finally able
to form cold gas giants with a combination of small planetesimals (5 km in
radius, violating the second constraint above) and very weak turbulence
($\gamma=10^{-5}$, violating the third constraint above). We conclude that the
formation of cold gas giants in the planetesimal accretion scenario requires not
only an extremely massive planetesimal population, but also a reduced
planetesimal size and a reduced turbulent stirring compared to the nominal
values. Our conclusions thus agree with those of \cite{Fortier+etal2013}, who
also found that the inclinations and sizes of the planetesimal population are
key parameters in setting the core growth rate.

Our conclusions about the inability for planetesimal accretion to form cold gas
giants under nominal conditions are in line with many studies that have used
$N$-body approaches to core growth by planetesimal accretion, both ignoring
migration \citep{Thommes+etal2003,Levison+etal2010} and including migration
\citep{ColemanNelson2014}. We have made many assumptions and simplifications to
make the problem tractable. Our most dubious assumption is maybe that there is
no interaction between the growing protoplanets. However, this approach allowed
us to study the physics of simultaneous growth and migration separately. We also
ignored the viscous heating of the protoplanetary disc and potential zones of
outwards migration in the inner parts of the protoplanetary disc
\citep{Cossou+etal2014}.

It is possible that we miss one or more important piece of physics in the
problem of core accretion with planetesimals. \cite{Capobianco+etal2011}
followed up on a finding from \cite{Levison+etal2010} showing that some of their
simulations starting with four massive protoplanets showed rapid outwards
migration of the protoplanets by planetesimal scattering and rapid growth to
cores of 10 Earth masses. \cite{ColemanNelson2016} instead considered the
possibility that the outer regions of the protoplanetary disc are filled with
sharp pressure bumps able to concentrate protoplanets and planetesimals and
stop migration to the inner disc. It is also possible that planetesimal
accretion in nature plays only a minor role for core accretion.  Alternatives to
the core accretion scenario include the direct collapse of the gas to form gas
giants \citep{Helled+etal2014} and the tidal downsizing of massive gas clumps to
form super-Earths \citep{Nayakshin2017}.

The pebble accretion scenario has also been proposed as an alternative to
planetesimal accretion \citep{JohansenLambrechts2017,Ormel2017}. The high
accretion rates of pebbles have been demonstrated in $N$-body simulations to
naturally form cold gas giants, regardless of whether migration is excluded
\citep{Levison+etal2015} or included
\citep{Izidoro+etal2019,Bitsch+etal2019}.  Statistical approaches including
pebble accretion and migration have found that pebble accretion can drive the
formation of cold gas giants in nominal protoplanetary disc models and for
large (drift-limited) and small (constant Stokes number ${\rm St}=0.01$) pebbles
\citep{Bitsch+etal2015,Johansen+etal2019}, provided that the turbulence is weak
enough to allow pebbles to sediment and form a relatively thin mid-plane layer
with a width of at most 10\% of the gas scale height. We present in Appendix A
growth maps for pebble accretion demonstrating that it is possible to form cold
gas giants by pebble accretion considering a conservative pebble growth model
where the pebble sizes are limited by fragmentation and bouncing. The area
of parameter space where cold gas giants emerge is nevertheless a strong
function of the turbulent diffusion coefficient and the metallicity of the
protoplanetary disc. This critical behaviour with metallicity is in good
agreement with exoplanet surveys that find an increasing occurrence rate of gas
giants beyond solar metallicity
\citep{Santos+etal2004,Johnson+etal2010,Buchhave+etal2012}.

Protoplanetary discs may very well have high masses in the earliest phases of
star formation \citep{Tychoniec+etal2018}, as envisioned in our models with a
massive primordial planetesimal population. However, while the formation of
1,000 Earth masses of planetesimals is questionable, the growth from the dust to
pebbles should be a natural process in these protoplanetary discs. The pebbles
that are not captured by the growing protoplanets would then simply be accreted
onto the star together with the gas.

Part of the drifting pebbles may also be converted into planetesimals, for example,\ at
the water-ice line \citep{SchoonenbergOrmel2017,DrazkowskaAlibert2017}.
\cite{Lenz+etal2019} showed that the trapping of drifting pebbles in vortices
and pressure bumps could form approximately 100 Earth masses of planetesimals,
far less than the 1,000 Earth masses needed to drive core growth by planetesimal
accretion. \cite{DrazkowskaDullemond2018}, on the other hand, found that several
hundred Earth masses of planetesimals can form from pebble pile-ups in a
relatively narrow region beyond the water-ice line. A large fraction of
this population of planetesimals ($\sim$10\%) would nevertheless have been
pushed into the asteroid belt by the migrating proto-Jupiter
\citep{RaymondIzidoro2017,Pirani+etal2019}, indicating that planetesimal
formation at the water-ice line cannot have been as efficient in the solar
system as found in \cite{DrazkowskaDullemond2018}. Photoevaporation provides an
additional route to convert hundreds of Earth masses of pebbles into planetesimals
out to 1,000 AU from the star \citep{Carrera+etal2017}, but such
populations of approximately 0.1-1 Earth masses per AU are not in
conflict with solar system evidence on the mass of the primordial planetesimal
population.

While pebble drift is an advantage because surplus planetary building
material is lost, the rapid loss of the pebble population could be detrimental to core
growth by pebble accretion. \cite{Johansen+etal2019} nevertheless demonstrated
that small pebbles, with Stokes numbers of about 0.01, continue to drift through
the disc for more than a million years if the initial protoplanetary disc is
large (of approximately 100 AU at the stage when the stellar accretion rate is
$10^{-7}\,M_\odot\,{\rm yr}^{-1}$). These small pebbles are brought along with
the viscous expansion of the outer regions of the protoplanetary disc and
subsequently drizzle in from large distances. Some protoplanetary discs observed
in recent ALMA surveys appear to be low mass and compact
\citep{Ansdell+etal2017}; such discs may have started out small and
rapidly lost most of their pebbles to radial drift. \cite{Manara+etal2018}
proposed that the low total mass of solids in relatively old protoplanetary
discs implies that planet formation was largely completed within the first million
years or sooner; early growth like this may again be explained in the pebble
accretion scenario given the potentially very high pebble mass fluxes at the
earliest stages of protoplanetary disc evolution
\citep{TanakaTsukamoto2019,Johansen+etal2019}.

As a final point, we advocate here using the giant planets in the solar system,
combined with constraints from the solar system planetesimal populations and
the observed turbulence levels in protoplanetary discs, as a benchmark problem
for planet formation models. If a planet formation theory fails to form the
solar system giants, given these constraints, then its
success at forming any exoplanetary systems is doubtful as well. 

\begin{acknowledgements}

AJ thanks the Swedish Research Council (grant 2018-04867), the Knut and Alice
Wallenberg Foundation (grants 2012.0150, 2014.0017) and the European Research
Council (ERC Consolidator Grant 724687-PLANETESYS) for research support. BB
thanks the European Research Council (ERC Starting Grant 757448-PAMDORA) for
their financial support. The authors would like to thank the anonymous referee
for helpful advice that helped improve the original manuscript.

\end{acknowledgements}

\appendix

\section{Pebble-driven planet formation model}

In this appendix we use a similar approach as in the main paper to perform
simulations of planet formation through pebble accretion, under constraints set by
observations of protoplanetary discs and by dust coagulation experiments. Our
approach to calculating the pebble accretion rate is described in detail in
\cite{Johansen+etal2019}.  We used the same protoplanetary disc model as for
planetesimal accretion, with the accretion rate starting at
$10^{-6}\,M_\odot\,{\rm yr}^{-1}$ and ending at $10^{-8}\,M_\odot\,{\rm
yr}^{-1}$ at disc dissipation after 3 Myr. The instantaneous pebble mass flux
was fixed at a constant ratio $\xi$ of the gas mass flux onto the star. The
pebble flux was assumed to be constant throughout the protoplanetary disc at any given
time, until the outer edge of the disc where the flux is damped with the same
exponential factor as the gas column density in the viscous accretion disc
evolution equations. A constant flux ratio like this could result from slow pebble
growth in the outer regions of the protoplanetary discs, so that pebbles grow
only from the dust transported inwards by the viscous gas flow at the outer edge
of the disc \citep[this model was explored in an appendix
of][]{Johansen+etal2019}. An important difference compared to the planetesimal
accretion models presented in the main paper is that the pebble column density
is set relative to the instantaneous gas mass flux onto the star, while the
planetesimals were assumed to form in a massive population at $t=0$.

In the spirit of choosing the free parameters according to constraints, we
considered pebble sizes based on coagulation experiments. We therefore set the
pebble size at a given distance from the star equal to the minimum of the size
obtained by fragmentation-limited pebble growth \citep{Birnstiel+etal2012} and
by bouncing-limited pebble growth \citep{Zsom+etal2010}. For
fragmentation-limited pebble growth we set the pebble Stokes number according to
\begin{equation}
  {\rm St} = \frac{1}{3} \delta^{-1} \left( \frac{v_{\rm f}}{c_{\rm s}}
  \right)^2 \, .
\end{equation}
This yields a turbulent collision speed equal to the  critical speed for
fragmentation, $v_{\rm f}$. The turbulent viscosity $\alpha_{\rm
v}$ sets the collision speed in reality, but we assumed here that the turbulent
viscosity is equal to the turbulent diffusion coefficient $\delta$. For the
bouncing-limited pebble growth we simply set the pebble radius equal to 1 mm.
\begin{figure}
  \begin{center}
    \includegraphics[width=\linewidth]{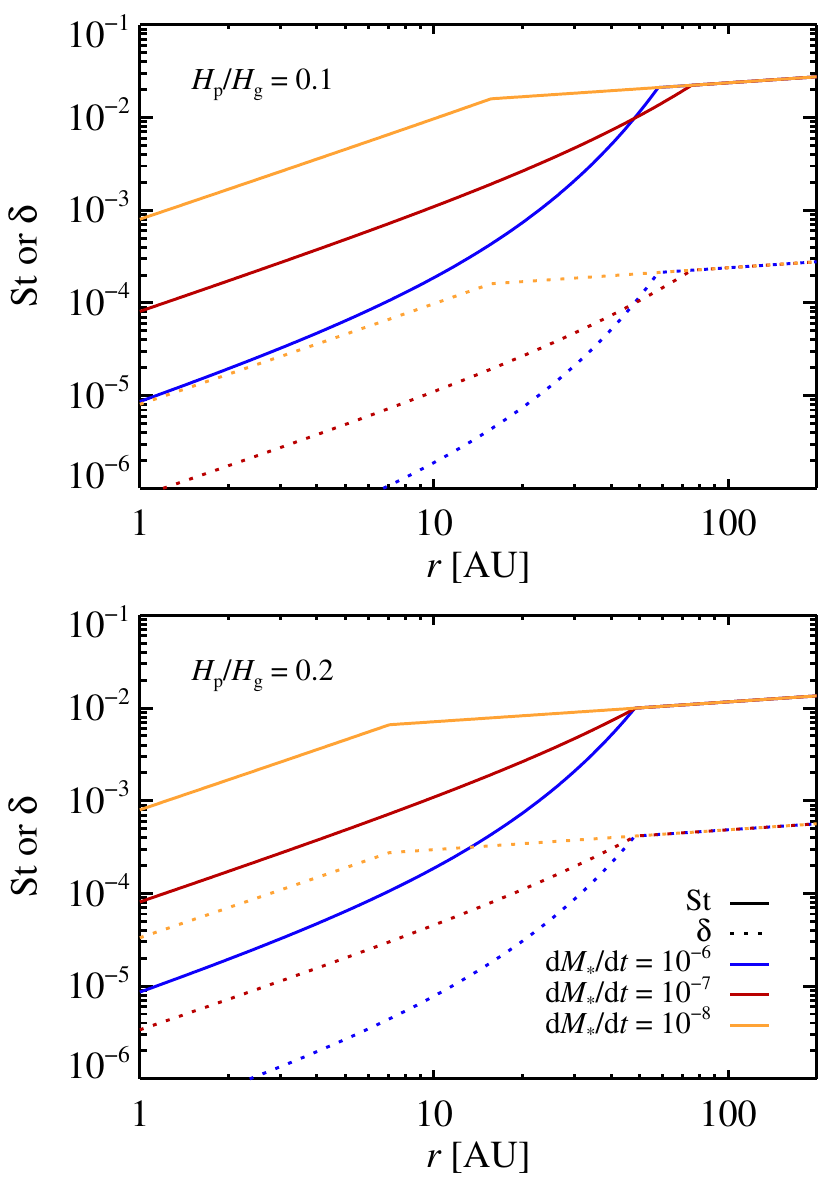}
    \caption{Stokes number (solid lines) and turbulent diffusion
    coefficient (dotted lines) for pebble growth limited by fragmentation and
    bouncing, shown for three different stages of protoplanetary disc evolution.
    The ratio of the scale height of the pebbles to the scale height of the gas
    is $H_{\rm p}/H_{\rm g}=0.1$ in the top plot and $H_{\rm p}/H_{\rm
    g}=0.2$ in the bottom plot. The outer regions of the protoplanetary disc
    have pebble sizes set by fragmentation, yielding a relatively constant
    Stokes number and diffusion coefficient. The bouncing-limited growth closer
    to the star gives very low values of the Stokes number and turbulent
    diffusion coefficient. Overall, the approximately millimeter-sized pebbles
    obtained in our model agrees with the pebble sizes inferred for the disc
    around the young star HL Tau \citep{Carrasco-Gonzalez+etal2019}.}
    \label{f:St_delta_r}
  \end{center}
\end{figure}

The turbulent diffusion coefficient was estimated by \cite{Pinte+etal2016} for
the protoplanetary disc around HL Tau from fitting synthetic images to the
observations. The best fit was found for a pebble scale height of 1 AU at a
distance of 100 AU from the star. At this distance the passively irradiated disc
model yields an aspect ratio of $H/r \approx 0.2$, assuming a central mass of
0.5 $M_\odot$ \citep{vanderMarel+etal2019} and a temperature of 47 K at 100 AU
\citep{Carrasco-Gonzalez+etal2019}. Hence the dust layer thickness relative to
the gas scale height is $H_{\rm p}/H_{\rm g} \approx 0.05$. This ratio arises
physically entirely from the Stokes number and the turbulent diffusion
coefficient through the diffusion-sedimentation equilibrium
\begin{equation}
  \frac{H_{\rm p}}{H_{\rm g}} = \sqrt{\frac{\delta}{{\rm St}+\delta}} \, .
  \label{eq:HpHg}
\end{equation}
The analysis of \cite{Pinte+etal2016} yielded a turbulent diffusion coefficient
slightly higher than $10^{-4}$ at 100 AU. \cite{MuldersDominik2012} estimated
similar values of turbulence by analysing the SED of protoplanetary discs in
a range of stellar masses. These values are estimated for the outer regions of
protoplanetary discs. \cite{Zhang+etal2018} proposed that a single planet of
mass $q=10^{-4}$ relative to the central star might be responsible for the
two bright rings in the protoplanetary disc around the young star AS 209 if the
turbulent viscosity is as low as $10^{-5}$ closer to the star.
\begin{figure*}
  \begin{center}
    \includegraphics[width=\linewidth]{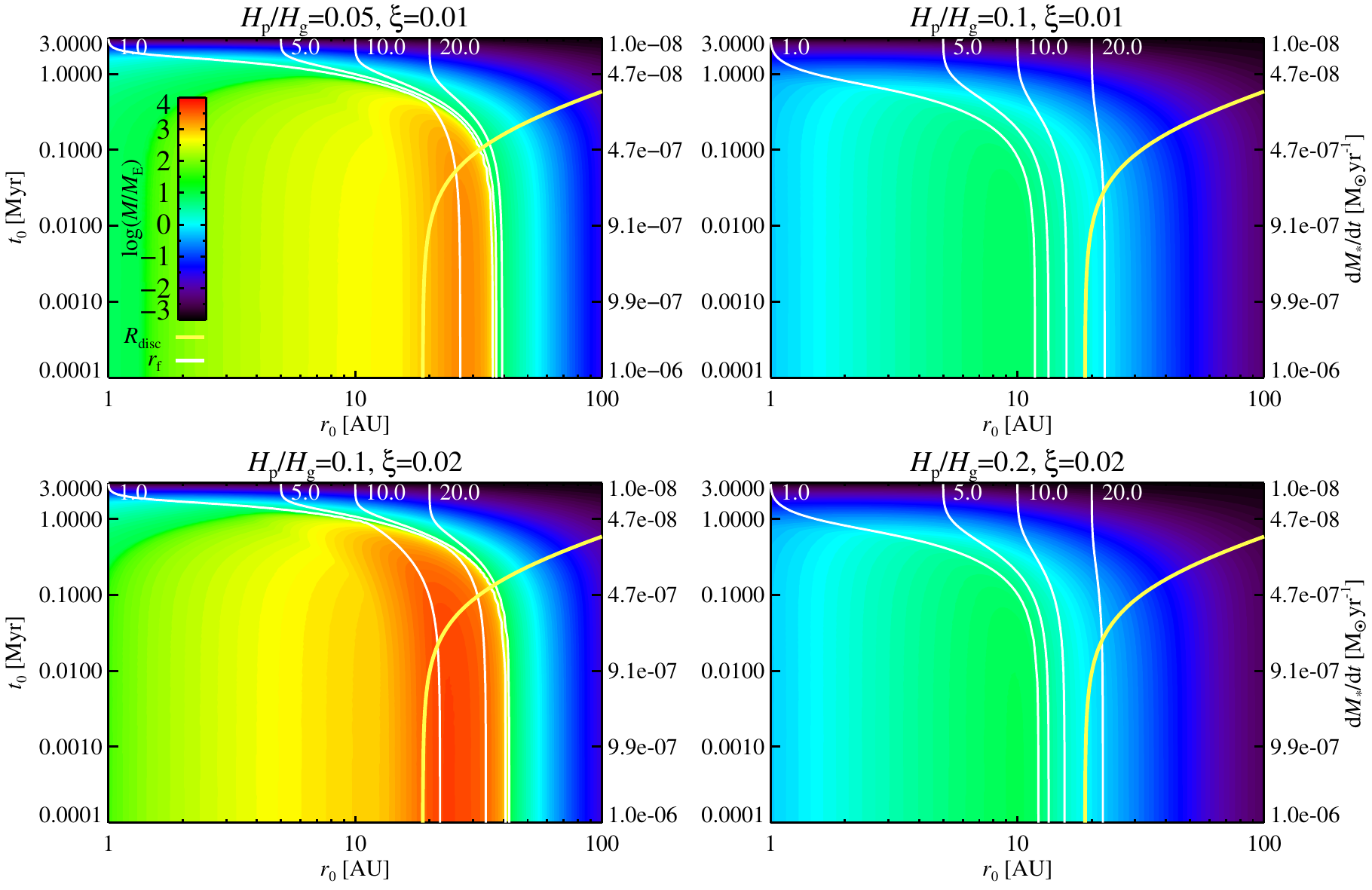}
  \end{center}
  \caption{Growth maps of protoplanets accreting pebbles and gas for four
  combinations of the relative pebble mid-plane layer thickness, $H_{\rm
  p}/H_{\rm g}$, and the pebble flux relative to the gas, $\xi$. Cold gas giants
  form readily from seeds starting in the outer parts of the protoplanetary disc
  when the mid-plane layer is very thin (top left plot) or the metallicity is
  higher than the solar value (bottom left plot). However, changing any of
  these parameters by a factor two increases the pebble accretion timescales
  enough to quench the formation of gas giants (top right and bottom right
  plots). Protoplanets only grow to super-Earth masses in these cases.}
  \label{f:growth_map_pebbles}
\end{figure*}

Knowing the scale height of the mid-plane layer and assuming
fragmentation-limited pebble growth yields a combined solution for the Stokes
number and the turbulent diffusion coefficient,
\begin{eqnarray}
  {\rm St} &=& \frac{1}{\sqrt{3}} \frac{v_{\rm f}}{c_{\rm s}}
  \sqrt{1/(H_{\rm p}/H_{\rm g})^2-1} \, , \\
  \delta &=& \frac{1}{\sqrt{3}} \frac{v_{\rm f}}{c_{\rm s}}
  \frac{1}{\sqrt{1/(H_{\rm p}/H_{\rm g})^2-1}} \, .
\end{eqnarray}
For the regions where growth is bouncing limited, we calculated the Stokes number
of the mm-sized pebbles stuck at the bouncing barrier and set the turbulent
diffusion coefficient that satisfies equation (\ref{eq:HpHg}) through
\begin{equation}
  \delta = \frac{\rm St}{1/(H_{\rm p}/H_{\rm g})^2-1} \, .
\end{equation}
In Figure \ref{f:St_delta_r} we show the Stokes number and turbulent diffusion
coefficient resulting from the pebble growth model. We considered two values
for the ratio of the pebble scale height to the gas scale height, $H_{\rm
p}/H_{\rm g}=0.1$ and $H_{\rm p}/H_{\rm g}=0.2$. Fragmentation sets the pebble
size in the outer regions of the protoplanetary disc. Fragmentation-limited
growth generally yields ${\rm St} \sim 0.01$ and $\delta \sim 0.0001$,
independent of the temporal evolution of the gas. These values are very similar
to the assumption of constant ${\rm St}=0.01$ and constant $\delta=0.0001$ of
\cite{Johansen+etal2019}. The growth becomes bouncing limited closer to the
star. The Stokes number of the millimeter-sized pebbles drops to very low
values, ${\rm St} \sim 10^{-5}-10^{-3}$. The accretion of such small pebbles by
pebble accretion is still poorly understood because their short gas coupling
timescales make them sensitive to recycling flows in the Hill sphere and
to convection \citep{Popovas+etal2018,Popovas+etal2019,RosenthalMurray-Clay2019}.
The planetary growth tracks that we obtain here within a few 10 AU should
therefore be interpreted with some caution; we focus in our analysis here on the
formation of cold gas giants farther from the star where the Stokes number is
higher.

We show growth maps for protoplanets accreting pebbles and gas in Figure
\ref{f:growth_map_pebbles} for four combinations of the relative thickness of
the pebble mid-plane layer $H_{\rm p}/H_{\rm g}$ and the pebble flux relative to
the gas flux $\xi$. For $H_{\rm p}/H_{\rm g}=0.05$, corresponding to the value
inferred for HL Tau \citep{Pinte+etal2016}, and $\xi=0.01$ cold gas giants
emerge from protoplanets that start to accrete pebbles between 30 and 35 AU (top
left panel). Protoplanets forming in the 10--30 AU region migrate enough to become gas giants on warm and hot orbits instead. Doubling the scale-height
ratio to $H_{\rm p}/H_{\rm g}=0.1$ (top right panel) quenches the formation of
giant planets, with protoplanets now only growing to super-Earth masses. At
twice the metallicity (bottom left panel), the cold gas giants reemerge, growing
even to super-Jupiter masses. Again doubling the relative scale height to
$H_{\rm p}/H_{\rm g}=0.2$ reverts the situation to forming only super-Earths
(bottom left panel). This transition from super-Earth formation to gas-giant
formation at around solar metallicity is consistent with exoplanet surveys
\citep{Buchhave+etal2012}. \cite{Johansen+etal2019} further demonstrated that
the parameter space for the formation of cold gas giants becomes much smaller
for initially smaller protoplanetary discs whose accretion rates drop faster.
The formation of gas giants in cold orbits is hard, even for pebble accretion.
The fact that the transition from super-Earths to gas giants occurs around
solar metallicity can nevertheless be seen as a major agreement between theory
and observations.

\end{document}